\documentclass[a4paper,11pt]{article}
\usepackage{pos}
\usepackage{graphicx}

\newcommand \hmu {\hat{\mu}}

\title{An overview of the QCD phase diagram at finite $T$ and $\mu$}

\author*[a,b]{Jana N. Guenther}

\affiliation[a]{CPT, Aix-Marseille Université, 163 Avenue de Luminy, 13009 Marseille, France}

\affiliation[b]{Theoretical particle physics, Bergische Universität Wuppertal, Gaußstraße 20, 42119 Wuppertal}

\emailAdd{jguenther@uni-wuppertal.de}

\abstract{In recent years there has been much progress on the investigation of the QCD phase diagram with lattice QCD. This talk will focus on the developments in the last few years. Especially the addition of external influences and extended ranges of $T$ and $\mu$ yield an increasing number of interesting results, a subset of which will be discussed. Many of these conditions are important for the understanding of both the QCD transition in the early universe and heavy ion collision experiments which are conducted for example at the LHC and RHIC. This offers many exciting opportunities for comparisons between theory and experiment.}

\FullConference{
 The 38th International Symposium on Lattice Field Theory, LATTICE2021
  26th-30th July, 2021
  Zoom/Gather@Massachusetts Institute of Technology
}

\begin{document}
\maketitle

\section{Introduction}
\label{sec:Introduction}

\begin{figure}
 \centering
 \includegraphics[width=0.6\textwidth]{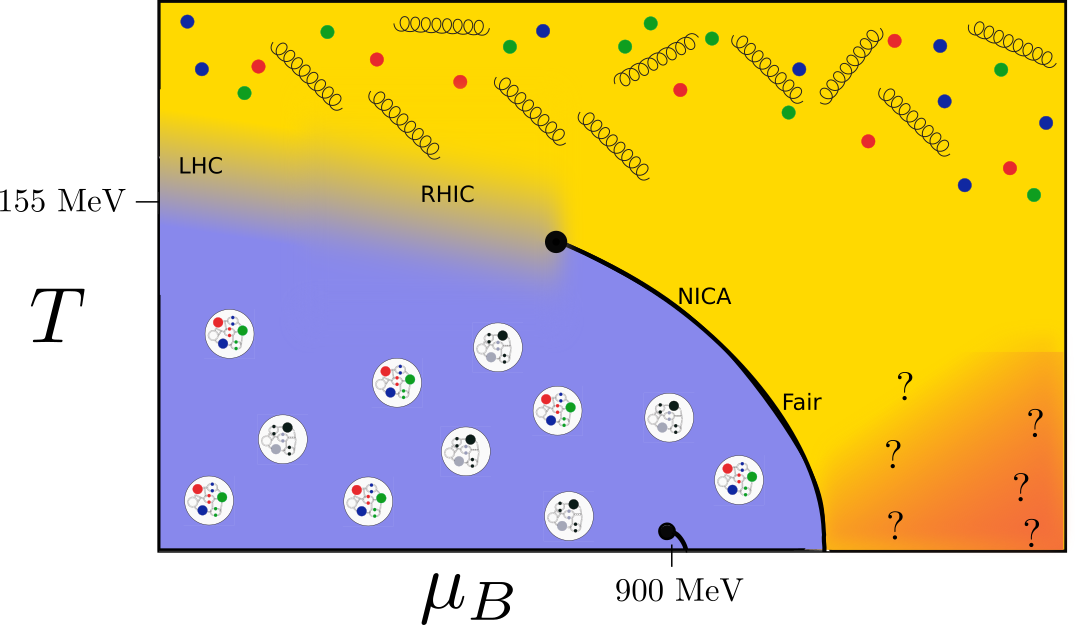}
 \caption{A schematic view on the $T$-$\mu_B$-plane of the QCD phase diagram. \label{fig:phaseDiagram}}
\end{figure}

The behaviour of QCD matter under different influences has been an active research topic for many years. The effects of different temperatures and densities are summerized in the $T$-$\mu$-plane of the QCD phasediagram which is schematicly displayed in figure~\ref{fig:phaseDiagram}. The investigation of its structure has seen segnificant progress in recent years broth in experiments and theory. 

Most experimental insights are gained from heavy ion collisions, most prominently done with gold or lead at the LCH and RHIC. Here two beams of heavy ions are collided at relativistic velocities, forming an out of equilibrium state, the so called glasma, which can be described as a color glass condensate (see for example Ref.~\cite{Gelis:2010nm,Gelis:2012ri}). Further fragmentation into quarks and gluons lead to the quark gluon plasma. This is a state of deconfined quarks and gluons that exhibits similarities with a strongly interacting fluid and is therefore often treated in the framework of relativisic hydrodynamics. After its formation the quark gluon plasma cools down again while expanding. When the quarks and gluons, which were in a deconfined state in the quark gluon plasma, recombine to colour-neutral hadrons, the chemical abundance of the different hadron species is fixed. This is called chemical freezout and it is assumed to take place at a similar temperature as the QCD transition. Even after the chemical freezout, the hadrons still can exchange momentum and energy. The time when this exchange comes to a stop is called kinetic feezeout.

On the theory side, lattice QCD is an obvious tool to investigate the phase diagram. It solves QCD with controllable errors, which allows for reliable predictions. Results from QCD thermodynamics in thermal equilibrium can be obtained with high precision for vanishing chemical potential. However, the investigation of finite densities is complicated by the infamous sign problem. Other non-perturbative methods like Dyson-Schwinger-Equations or Functional Renmormalization Groups do not encounter the sign problem, but fail to determine a reliable error.

Since lattice QCD simulates quantities in thermal equilibrium, the question at which states the quark gluon plasma (or the hadrons) is thermalized is very important for comparisons between experimental and lattice QCD results. The state of the glasma does not thermalize and is therefore difficult to investigate with lattice QCD.

This proceedings will focus on the review of recent progress obtained from lattice QCD. A focus will be on results with physical parameters at low finite $\mu_B$ by extrapolations from zero or imaginary $\mu_B$, which allow for comparisons with experimental results. Other reviews about lattice QCD results that can not be discussed here are for example Refs.~\cite{Philipsen:2019rjq, Guenther:2020jwe}.

\subsection{Extrapolation to small finite chemical potential\label{sec:extrap}}

Since at $\mu_B=0$ the transition is a crossover 
(Ref.~\cite{Aoki:2006we,Aoki:2006br,Aoki:2009sc,Borsanyi:2010bp,Bhattacharya:2014ara,Bazavov:2011nk}), 
some observables can be described by an analytic function in the vicinity of zero. This fact can be exploited, by using results at imaginary or zero chemical potential. A very common technique for extrapolation is the Taylor method. Within that method, the pressure is parameterized as 
\begin{equation}
   \frac{p}{T^4}=\sum_{j=0}^\infty \sum_{k=0}^\infty \frac{1}{j! k!} \chi^{BS}_{jk} \hat \mu_B^j \hat \mu_S^k
   \label{eqn:taylor}
  \end{equation}
  with $\hat \mu=\frac{\mu}{T}$. The $ \chi^{BS}_{jk}$ can be measured at zero $\mu$. Another expansion from which results will be discussed in this work, is the fugacity expansion or sector method. Here the parameterization 
  \begin{equation}
   \frac{p}{T^4}=\sum_{j=0}^\infty \sum_{k=0}^\infty P^{BS}_{jk} \cosh( j\hat \mu_B  -k\hat \mu_S)
   \label{eqn:fugacity}
  \end{equation}
 is used. The expansion coefficients $P^{BS}_{jk}$ have to be determined either from the $\chi^{BS}_{jk}$ or from imaginary chemical potential. In general, for simulations with imaginary chemical potential a wide variety of fit functions are possible to describe the data at $\mu^2<0$. The choice of fit function is usually guided by the fit quality, sometimes complemented by known physical insights. An advantage of the fugacity expansion is its rapid convergence in the hadronic phase and the added information on the particle contend from different sectors. On the other hand, the Taylor expansion convergences rapidly in the high temperature limit.
 
 \subsection{Simulations at finite chemical potential}
 Since the extrapolation methods discussed in section~\ref{sec:extrap} are only feasible for small chemical potential and for an analytic transition other techniques are required to reach out further into the phase diagram. There have been many ideas around in the Lattice community how to facilitate simulations despite the sign problem, for example reweighting techniques \cite{Barbour:1997ej,Fodor:2001au,Fodor:2001pe,Csikor:2002ic}, density of state methods \cite{Fodor:2007vv,Alexandru:2014hga}, using the canonical ensemble \cite{Alexandru:2005ix,Kratochvila:2005mk,Ejiri:2008xt}, formulations with dual variables \cite{Gattringer:2014nxa} or Lefschetz thimbles \cite{Scorzato:2015qts, Alexandru:2015xva}. Two methods for which I will briefly discuss some resent results are Complex Langevin and Sign reweighting. While non of the available results are at physical quark masses and continuum extrapolated, and might still have some other caviates that prevent them from giving final answers on the phase diagram, impressive progress has been made recently.
 \paragraph{Complex Langevin}
 One way to simulate QCD with finite density, that has been intensely studied and made significant progress over recent years is to employ Complex Langevin equations. While here only a few recent result will be discussed, a more comprehensive overview can can be found in Ref.~\cite{Attanasio:2020spv}. Complex Langevin simulation extend the gauge group from $SU(3)$ to $SL(3,\mathbb{C})$. The non-compact nature of $SL(3,\mathbb{C})$ can lead to so called runaway configurations. To make sure the evolution stays close to the unitary manifold, which produces the correct result, gauge cooling (Refs.~\cite{Seiler:2012wz,Aarts:2013uxa}) can be employed. Other methods for improvement are the use of an adaptive step size for the numerical integration (Ref.~\cite{Aarts:2009dg}) and the addition of a force term to the evolution (Ref.~\cite{Attanasio:2018rtq}). The correctness of the result can be monitored by checking the fall-of of specific observables, which usually is required to meet a certain speed (Refs.~\cite{Nishimura:2015pba,Aarts:2017vrv,Nagata:2016vkn, Nagata:2018net, Scherzer:2018hid,Tsutsui:2019suq}).
 
 Recent results on the phase diagram with Complex Langevin simulations include the use of improved actions like the Symanzik improved gauge action and the comparison to results from the Tayor expansion method (see section~\ref{sec:extrap}) with was done in Ref.~\cite{Sexty:2019vqx} with four flavours of staggered quarks with pion masses between 500~MeV and 700~MeV on $16^3\times8$ lattices. 
 
 Also, for heavy pions with a mass of about 1.3~GeV and two flavors of Wilson fermions the transition temperature at finite $\mu$ was computed in (Ref.~\cite{Scherzer:2020kiu}). The determination of the transition temperature was obtained from the third order Binder cumulant of two different observables, both related to the Polyakov loop.
 
 In Ref.~\cite{Ito:2020mys,Namekawa:2021qtg,Tsutsui:2021bog} Complex Langevin simulations in small boxes with sizes of $8^3\times16$ and $16^3\times8$ with four flavors and a lattice spacing of $a^{-1}\approx4.7$~GeV were performed. For the dependence of the quark number on the chemcial potential a plateau was observed that the authors relate to the Fermi surface and color superconductivity.
 
\begin{figure}
 \centering
 \includegraphics[width=0.45\textwidth]{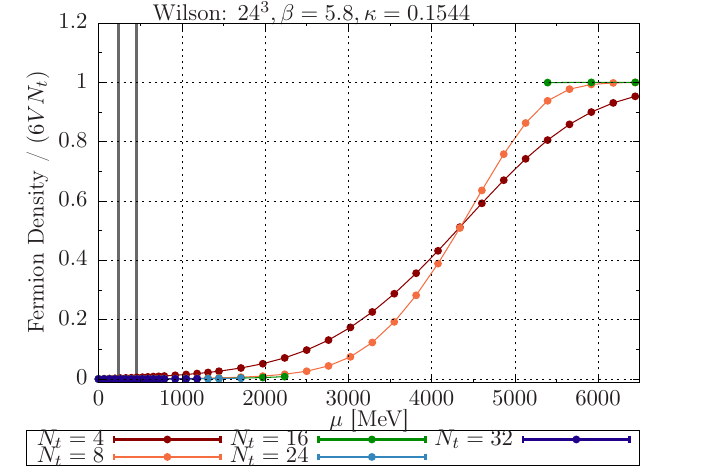}
 \hfill
 \includegraphics[width=0.45\textwidth]{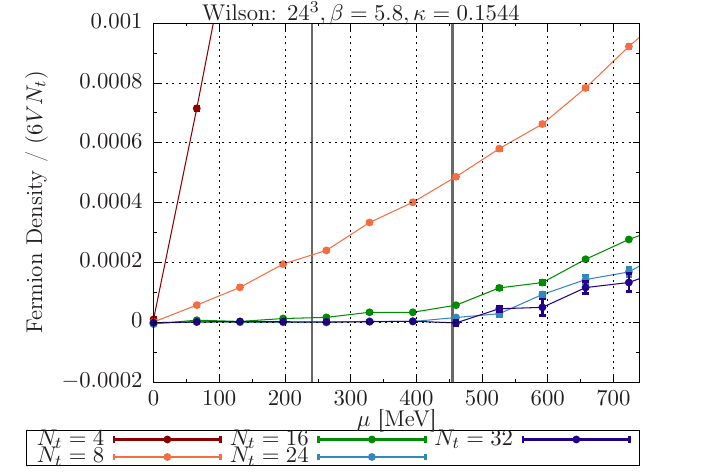}
 \caption{(Ref.~\cite{Attanasio:2021hyh}) The fermion density at finit chemcial potential from Complex Langevin simulations. Left: Overview over a large range of chemical potential. Right: Zoom in of the lower left corner of the right side.}
 \label{fig:CL}
\end{figure}

 Results for a large range of chemical potentials and several lattice spacings and volumes were presented at this year's lattice conference (Ref.~\cite{Attanasio:2021hyh}). Here two flavors of Wilson fermions ($m_\pi=550$~MeV) were used for simulations of chemical potentials up to 5~GeV. The results for fermion density normalized by a factor of $\frac{1}{6VN_t}$ are shown in figure~\ref{fig:CL}. They are shown for lattices with $N_s=24$ and five different values of $N_t$ between 4 and 32. For large chemical potential a saturation effect can be observed. For the simulations, adaptive step size scaling, gauge cooling and dynamic stabilisation were employed. The results still have to be extrapolated to a Langevin step size of zero.
 
 \paragraph{Reweighting}
 A common challenge when reweighing from zero to finite chemical potential is the overlap problem: The amount of configurations obtained by importance sampling that contain information on finite $\mu$ physics is prohibitively small because they are only in the tails of the probability distribution.
 To mitigate this overlap problem one can reweight from a theory where the reweighing factors are from a compact space. Two possible choices, for which results were presented at this conference (Ref.~\cite{Pasztor:2021ray}), are the phase quenched theory, where the reweighing factors are pure phases $e^{i\theta}$ and the sign quenched theory (Refs.~\cite{deForcrand:2002pa,Alexandru:2005ix,Giordano:2020roi}), which also has a weaker sign problem.
 New results from this reweighting techniques were presented in Refs.~\cite{Pasztor:2021ray,Borsanyi:2021hbk} for lattice with $N_t=4$ and 6.
 
\section{The transition temperature}

The crossover nature of the transition and it temperature has been determined since 2006 with a variety of observables (Ref.~\cite{Aoki:2006we,Aoki:2006br,Cheng:2006qk,Aoki:2009sc,Bazavov:2009zn,Borsanyi:2010bp,Bhattacharya:2014ara,Bazavov:2011nk}). More recently the transition temperature has been determined with increased precision in Ref.~\cite{Bazavov:2018mes} and Ref.~\cite{Borsanyi:2020fev}. 
Often different definitions yield consistent results within the available precision. However for an analytic transition this is not guaranteed in contrast to the situation of a phase transition. 

A high precision determination of the transition temperature at $\mu_B=0$ from five different observables was done in Ref.~\cite{Steinbrecher:2018phh}. All five definitions have the same continuum limit within the available precision. This yields a combined value of $T_c=(156\pm1.5)$~MeV.

A new definition as the peak of the chiral susceptibility as a function of the chiral condensate (instead of the more common definition as function of the temperature) was introduced in Ref.~\cite{Borsanyi:2020fev}. It allows for a more precise extraction of the transition temperature and, therefore, for an improvement in the extrapolation (see section~\ref{sec:finiteMu}). 

\begin{figure}
 \centering
 \includegraphics[width=0.5\textwidth]{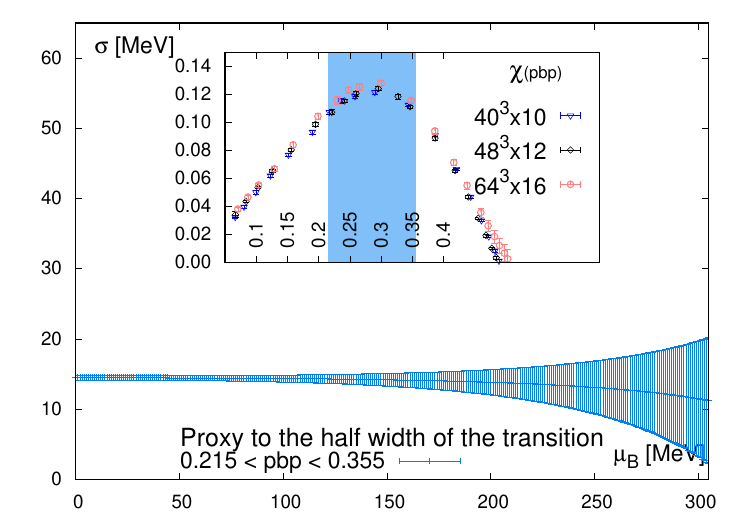}
 \caption{(Ref.~\cite{Borsanyi:2020fev}) Extrapolation of the half width of the transition to finite baryon chemical potential as defined in equation~(\ref{eqn:width}). The insert shows the chiral suceptibility $\chi$ as a function of the chiral condensate $\langle \overline{\psi} \psi \rangle$. The shaded region illustrates $\langle \overline{\psi} \psi \rangle_c \pm \Delta \langle \overline{\psi} \psi \rangle/2$.}
 \label{fig:width}
\end{figure}

By now the error on the transition temperature at $\mu_B=0$ is much smaller than the width $\sigma$ of the crossover. It has to be determined separately and is not encompassed by the error of the transition temperature. A possible definition, that has been used in Ref.~\cite{Borsanyi:2020fev}, is
\begin{equation}
 \langle \overline{\psi} \psi \rangle (T_c\pm \sigma/2) = \langle \overline{\psi} \psi \rangle_c \pm \Delta \langle \overline{\psi} \psi \rangle/2. \label{eqn:width}
\end{equation}
The result, extrapolated to the continuum from three lattice spacings, can be seen in figure~\ref{fig:width} up to $\mu_B=300$~MeV.

\subsection{Influence of many colors}
\begin{figure}
 \centering
 \includegraphics[width=0.32\textwidth]{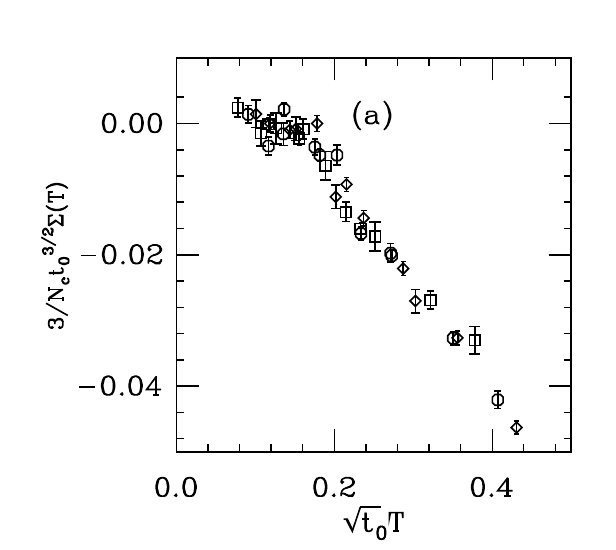}
 \hfill
 \includegraphics[width=0.32\textwidth]{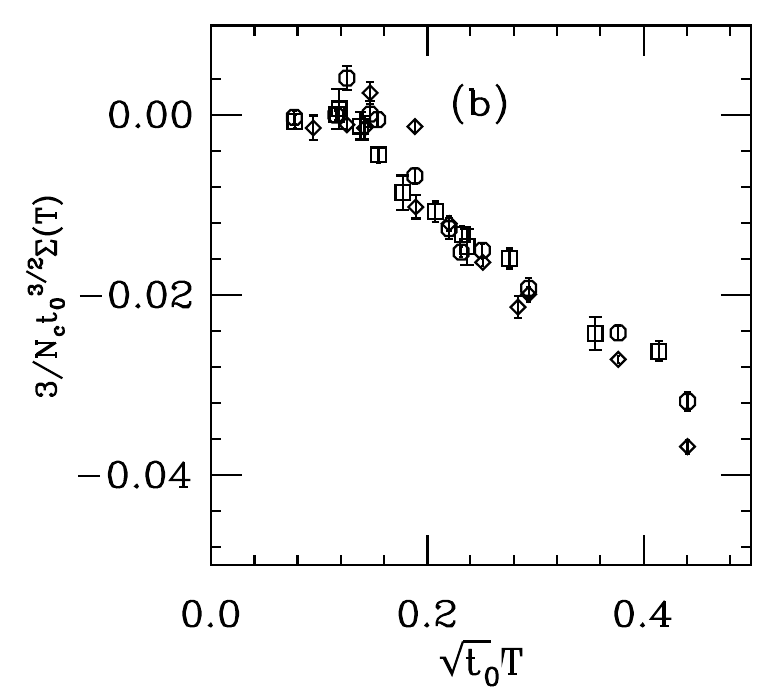}
 \hfill
 \includegraphics[width=0.32\textwidth]{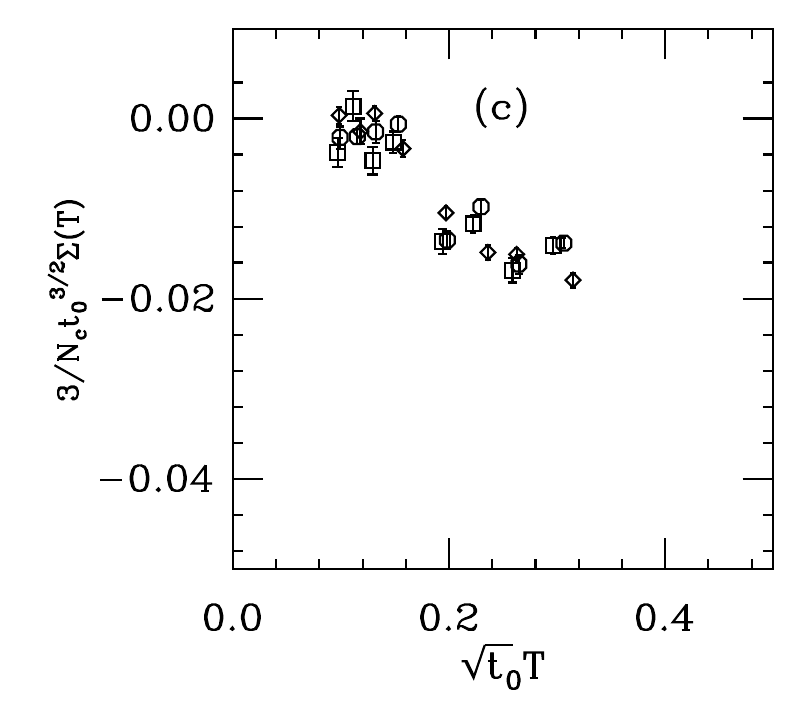}
 \caption{(Ref.~\cite{DeGrand:2021zjw}) The rescaled chiral condensate as a function of the temperature in units of $t_0$. The number of colors is shown by the shapes: Squares: $N_c=3$, octagons: $N_c=4$ and diamonds: $N_c = 5$. The three plots show different pseudo scalar masses:
(a)  $(m_{PS}/m_V)^2 \sim 0.63$;
(b) $(m_{PS}/m_V)^2 \sim 0.5$;
(c) $(m_{PS}/m_V)^2 \sim 0.25$.
}
\label{fig:color}
\end{figure}

On possible influence on the transition temperature is the number of colors. The large color limit is theoretically interesting as it simplifies certain aspects of QCD (see for example \cite{tHooft:1973alw,tHooft:1974pnl,Witten:1979kh}). Recent reviews for lattice QCD with $N_c>3$ can be found in Refs.~\cite{GarciaPerez:2020gnf,Hernandez:2020tbc}. At this conference the results of Ref.~\cite{DeGrand:2021zjw} where presented (Ref.~\cite{DeGrand:2021xha}). This work investigates $T_c$ with $N_f=2$ Wilson-Clover fermions for $N_c=3,4$ and 5. Figure~\ref{fig:color} shows the temperature dependence of the rescaled chiral condensate for three different pseudo scalar masses. After including the rescaling factor $\frac{3}{N_c}$, the chiral condensate is within the available precision independent of the number of colors. 

 \subsection{Quark masses and the Columbia plot\label{sec:Columbia}}
\begin{figure}
  \centering
  \includegraphics[width=0.4\textwidth]{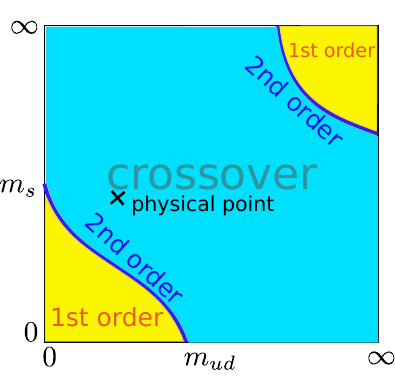}
  \hspace{1cm}
  \includegraphics[width=0.4\textwidth]{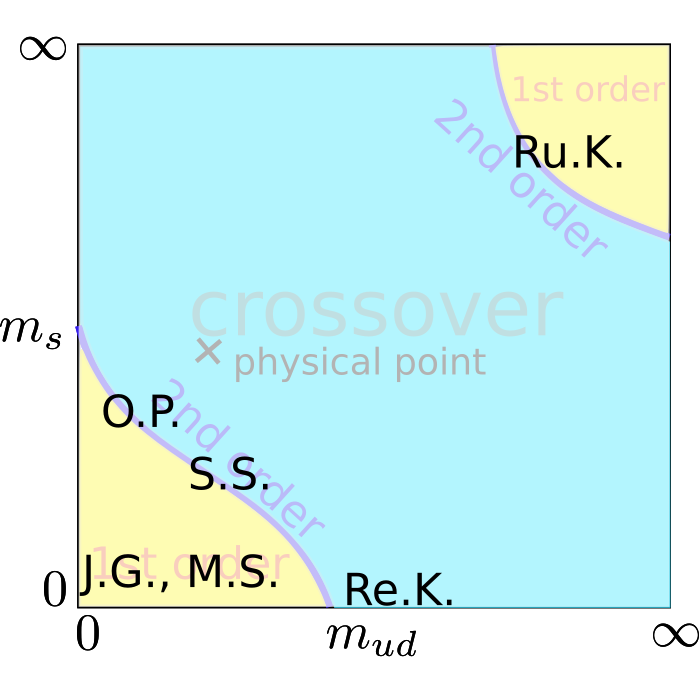}
  \caption{A schematic view of one possible version of the Columbia. On the left side the talks at this years's Lattice conference are noted as well (Refs~\cite{Cuteri:2020yke,Kara:2021btt,Clarke:2021ljn,Dini:2021hug,Philipsen:2021qji,Cuteri:2021ikv}).}
  \label{fig:Columbia}
 \end{figure}

 An important influence on the QCD transition temperature is its dependence on the light and strange quark masses. The Columbia plot (figure~\ref{fig:Columbia}) shows one possible scenario for the dependence of the order of the transition as a function of the quark masses for $N_f=2+1$. The infinite quark mass limit (the upper right corner) yields pure $SU(3)$-gauge theory with static quarks where the QCD transition is of first order. When the quark masses become finite, the first order transition is getting weaker until it becomes a second order transition. On the opposite corner of the Columbia plot (the lower left corner) the chiral limit for three flavours is expected to be a first order transition as well. Again, when quark masses become larger the transition weakens until it becomes a second order transition. The limiting second order lines for both corners are still under active investigation. Several results on the Columbia plot where presented at this years lattice conference, and are roughly depicted on the right side of figure~\ref{fig:Columbia}. However, a dedicated review of, especially the chiral limit is given in Ref.~\cite{ColumbiaReview,TalkColumbia} and not part of this proceedings.
 
 \subsection{Extrapolation to finite $\mu_B$\label{sec:finiteMu}}

\begin{figure}
 \centering
 \includegraphics[width=0.8\textwidth]{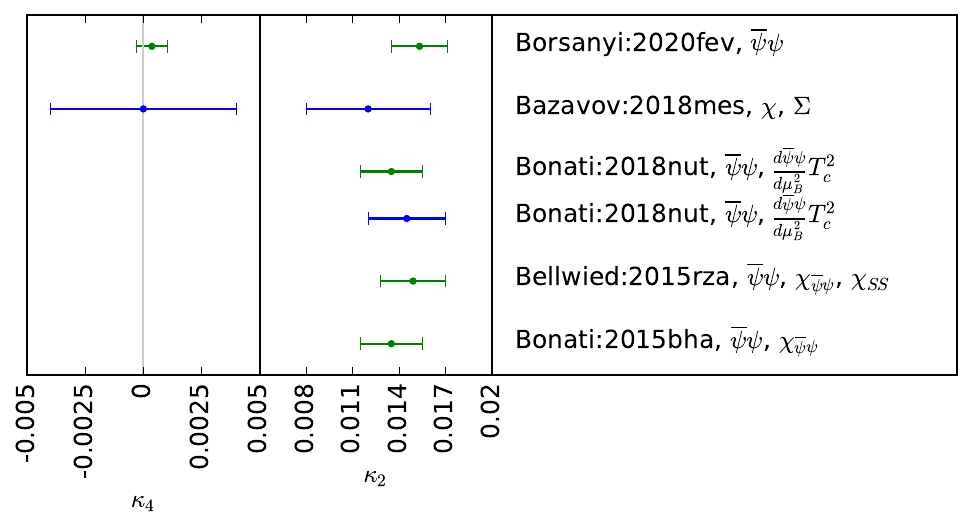}
 \caption{(Ref.~\cite{Borsanyi:2020fev}) Overview of the different determinations of the $\kappa_2$ and $\kappa_4$ coefficients as defined in equation~(\ref{eqn:kappaLambda}).  The values are taken from Ref.~\cite{Borsanyi:2020fev,Bazavov:2018mes,Bonati:2018nut, Bellwied:2015rza,Bonati:2015bha}. Green points correspond to determinations from imaginary chemical potential, while results shown in blue were obtained by the Taylor method.}
 \label{fig:kappaLambda}
\end{figure}

The behaviour of the transition temperature in the $\mu_B$-$T$-plane can be parameterized by the Taylor expansion as 
\begin{equation}
 \frac{T_c(\mu_B)}{T_c(0)} = 1 - \kappa_2 \left(\frac{\mu_B}{T_c} \right)^2  - \kappa_4 \left(\frac{\mu_B}{T_c} \right)^4   + \mathcal O(\mu_B^6).\label{eqn:kappaLambda}
\end{equation}
The odd powers of $\frac{\mu_B}{T}$ vanish due to the isospin symmetry. The coefficients $\kappa_2$ has been determined in several computations. The more recent ones from Refs.~\cite{Borsanyi:2020fev,Bazavov:2018mes,Bonati:2018nut,Bellwied:2015rza,Bonati:2015bha} are compared in  figure~\ref{fig:kappaLambda}. The green points were obtained from extrapolations with imaginary chemical potential while the blue points were obtained by the Taylor method. Both methods yield compatible results. In addition to $\kappa_2$ new results for $\kappa_4$ are available from Refs.~\cite{Bazavov:2018mes, Borsanyi:2020fev}. While it is clear that $\kappa_4 \ll \kappa_2$ the relative error of $\kappa_4$ is more than 100\% and the sign is therefore still undetermined.

\begin{figure}
 \centering
 \includegraphics[width=0.49\textwidth]{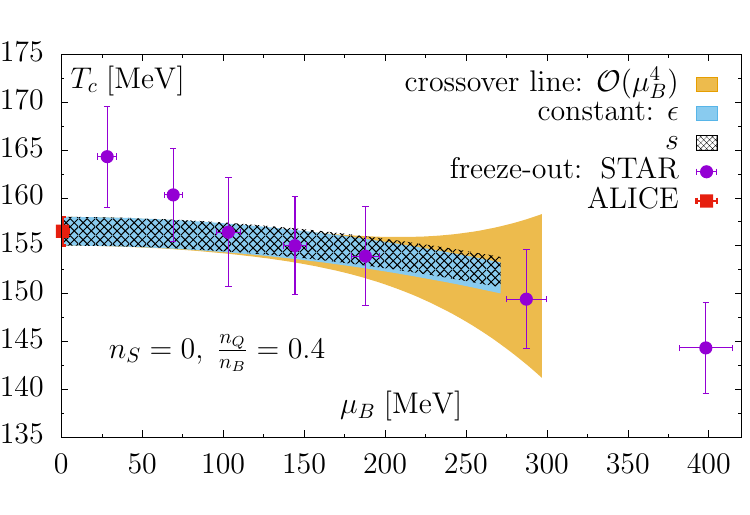}
 \hfill
 \includegraphics[width=0.49\textwidth]{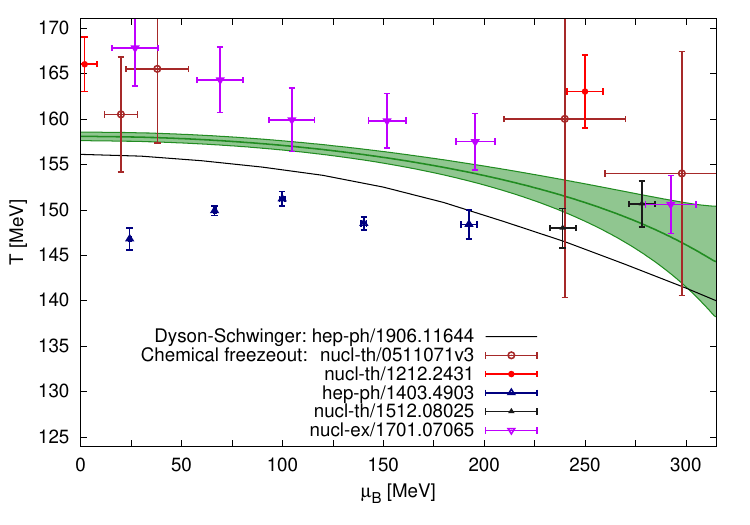}
 \caption{Left: (Ref.~\cite{Bazavov:2018mes}) The extrapolation of the transition from the Taylor method compared with the freeze out temperatures from Ref.~\cite{Andronic:2017pug,STAR:2017sal}, Right: (Ref.~\cite{Borsanyi:2020fev}) A comparison of the extrapolated transition temperature from simulations at imaginary chemical potential (green band) and Dyson-Schwinger-Equations (Ref.~\cite{Isserstedt:2019pgx}). In addition the freeze-out temperatures obtained from heavy ion collisions in Ref.~\cite{Andronic:2005yp,Becattini:2012xb,Alba:2014eba,Vovchenko:2015idt,Adamczyk:2017iwn} are shown.  \label{fig:TcExtrap}}
\end{figure}

The extrapolation of the transition temperature to finite $\mu_B$ for the choice of vanishing net strangeness ($n_S=0$) and $n_Q=0.4n_B$, to match the condition in heavy ion collisions from Refs.~\cite{Bazavov:2018mes, Borsanyi:2020fev} is shown in figure~\ref{fig:TcExtrap}. The results from Ref.~\cite{Bazavov:2018mes} (left side of figure~\ref{fig:TcExtrap}) rely on the Taylor expansion method. They were obtained with HISQ quarks and continuum extrapolated from three lattices with temporal extend $N_t=6,8,12$.

The results on the right hand side of figure~\ref{fig:TcExtrap} were obtained from simulations at imaginary chemical potential. They are continuum extrapolated from three lattices with sizes $40^3\times10$, $48^3\times12$ and $64^3\times16$ and obtained using stout smeard staggered quarks.. The extrapolation was done with two different functions,
\begin{equation}
 T_c=1 + \hmu_B^2 \left( a+ \frac{d}{N_t^2}\right) + \hmu_B^4 \left( b+ \frac{e}{N_t^2}\right) + \hmu_B^6 \left( c+ \frac{f}{N_t^2}\right)
\end{equation}
and
\begin{equation}
 T_c=\frac{1}{1 + \hmu_B^2 \left( a+ \frac{d}{N_t^2}\right) + \hmu_B^4 \left( b+ \frac{e}{N_t^2}\right) + \hmu_B^6 \left( c+ \frac{f}{N_t^2}\right)}
\end{equation}
to estimate the systematic error from the choice of extrapolation function.

As a comparison with the lattice result several data points for the freeze-out temperature (Refs.~\cite{Andronic:2005yp,Becattini:2012xb,Alba:2014eba,Vovchenko:2015idt,Adamczyk:2017iwn, Andronic:2017pug,STAR:2017sal} are shown. While the chemical freeze-out is not the same as the QCD transition, it is expected to occur at a similar temperature, which makes an comparison interesting. In a addition, on the right side, a result from Dyson-Schwinger-Equation calculations (Ref.~\cite{Isserstedt:2019pgx}) is shown. The curvature agrees well with the lattice result, while in that case the absolute value was set to a previous lattice value, which was determined by a different observable and with a larger error. Therefore the difference does not imply a contradiction between the two calculations.

\subsection{The influence of a magnetic field}
\begin{figure}
 \centering
 \includegraphics[width=0.9\textwidth]{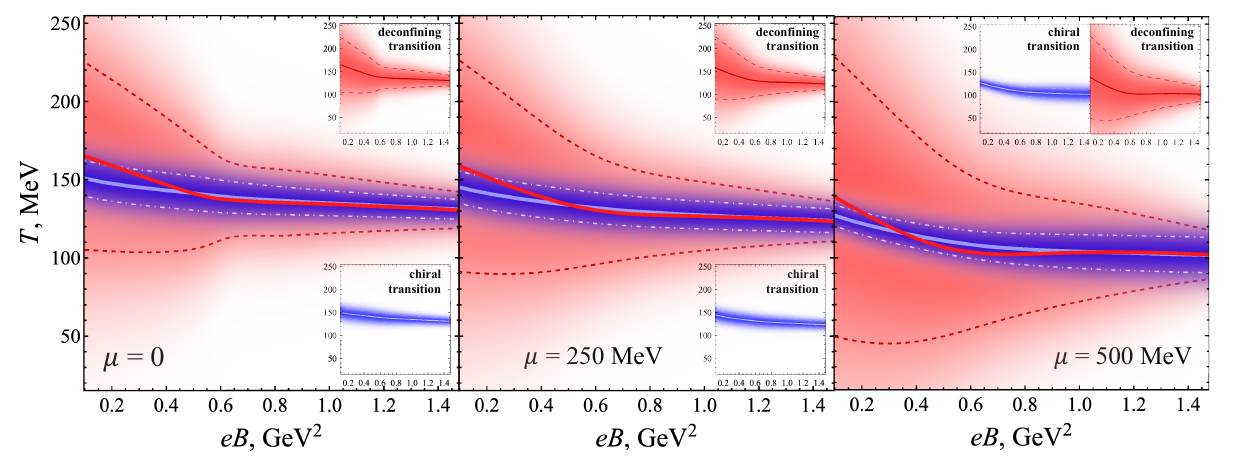}
 \caption{(Ref.~\cite{Braguta:2019yci}) The transition temperature, both for the chiral and deconfinement transition, as a function of the magnetic field, for three finite densities.}
 \label{fig:magnetic}
\end{figure}

When considering the situation in heavy ion colliders, another important influence on the transition temperature is the magnetic field (Refs.~\cite{Kharzeev:2007jp,Skokov:2009qp,Deng:2012pc}). In the last decade the simulation of QCD with a magnetic field on the lattice has been a very active field (e.g. Refs.~\cite{DElia:2010abb,Bali:2011qj,Bali:2012zg,Shovkovy:2012zn,Ilgenfritz:2013ara,Bornyakov:2013eya,Bali:2014kia,Endrodi:2019zrl,Tomiya:2019nym,Ding:2020hxw}). In several works, which were not yet continuum extrapolated, an increase of the transition temperature with the addition of a magnetic field was found. This agreed well with the expectation resulting from the so-called magnetic catalysis, which describes that at zero temperature the chiral symmetry breaking increases with the magnetic field. However, continuum extrapolated results find the opposite behaviour, the so called inverse magnetic catalysis. Studies with various pion masses (Refs.~\cite{DElia:2018xwo,Endrodi:2019zrl}) suggest that this effect is related to the deconfinemend rather than the chiral transition. The effects of a magnetic field remained an active topic at this year's lattice conference \cite{Braguta:2021ucr,Astrakhantsev:2021jpl,DElia:2021yvk,TalkMaio,Brandt:2021vez,Ding:2020hxw,TalkWang,Ding:2021cwv,TalkLi}. In addition to studying a magnetic field at zero or finite temperature, now also the combination of a magnetic field and a finite density is under active investigation (Refs.~\cite{Braguta:2019yci,Braguta:2021ucr,Astrakhantsev:2021jpl}). Figure~\ref{fig:magnetic} shows the transition temperature, both for the chiral and deconfinement transition, as a function of the magnetic field, for three chemical potentials ($\mu_B= 0~\mathrm{MeV}, 250~\mathrm{MeV},500~\mathrm{MeV}$). The results were obtained on $N_t=6$ lattices with staggered quarks. They show that the addition of a chemical potential further decreses the transition temperature, which matches the expectation from extrapolation to finite density without a magnetic field (see section~\ref{sec:finiteMu})

\section{Fluctuations}
Fluctuations are computed as the derivatives of the pressure with respect to various chemical potentials:
\begin{equation}
     \chi^{B,Q,S}_{i,j,k}= \frac{\partial^{i+j+k} (p/T^4)}{
(\partial \hat\mu_B)^i
(\partial \hat\mu_Q)^j
(\partial \hat\mu_S)^k
}\,, \ \hat\mu_i=\frac{\mu}{T}
\end{equation}

\begin{figure}
 \centering
 \includegraphics[width=0.5\textwidth]{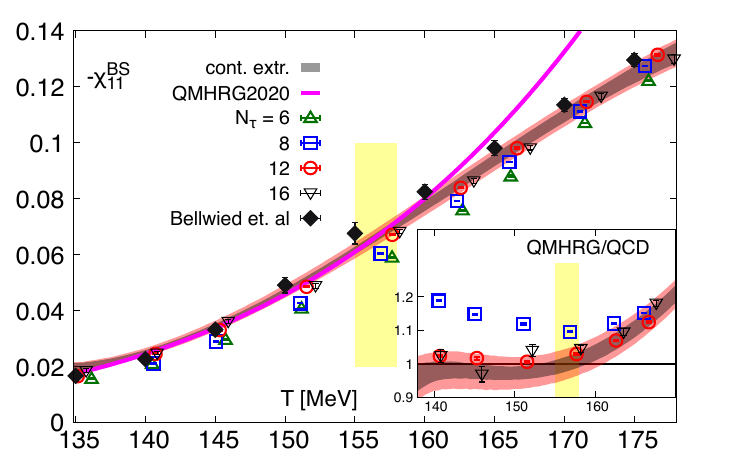}
 \caption{(Ref.~\cite{Bollweg:2021vqf}) New continuum extrapolated results ($N_t=6,8,12,16$) allow for detailed comparisons with various models.}
 \label{fig:chiBS}
\end{figure}

While fluctuations to various order have previously published for example in Ref.~\cite{Schmidt:2012ka, DElia:2016jqh,Bazavov:2017dus, Borsanyi:2018grb,Bazavov:2020bjn} ,now new continuum extrapolated results are available in Ref.~\cite{Bollweg:2021vqf}. These results are obtained by the Taylor method and continuum extrapolated from lattices with $N_t = 6,8,12$ and 16 with HISQ fermions.  The precision of these results is high enough to allow for a comparison to different models with detailed studies for example on inclusion or exclusion of various states in a Hadron Resonance Gas model, as shown figure~\ref{fig:chiBS}. To match the lattice results, for example for $\chi^{BS}_{11}$, it is necessary to add states from quark models to the list of resonances from the PDG \cite{ParticleDataGroup:2020ssz}.

\begin{figure}
 \centering
  \includegraphics[width=0.5\textwidth]{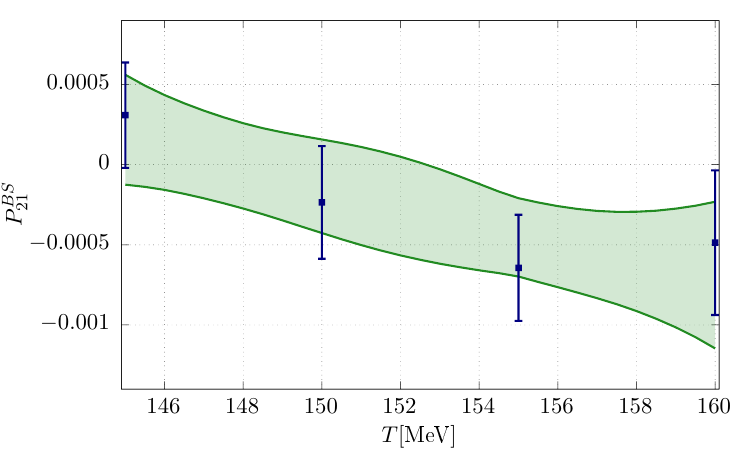}
  \caption{(Ref.~\cite{Bellwied:2021nrt}) Continuum estimate for the fugacity expansion coefficient (see equation~(\ref{eqn:fugacity})) from $N_t=8,10,12$ with stout smeared staggered fermions.  This coefficient includes contributions from $N-\Lambda$ and $N-\Sigma$ scattering. The negative trend indecates the presence of an repulsive interaction that cannot be described with the addition of more resonances.}
  \label{fig:PBS21}
\end{figure}

On the other hand in Refs~\cite{Bellwied:2021nrt,Bellwied:2021skc} results on the fugacity expansion coefficients (see equation~(\ref{eqn:fugacity})) from imaginary chemical potential are presented. Here the results are continuum estimates obtained with stout smeard staggered fermions on $N_t=8,10$ and 12 lattices. The analysis is based on a two dimensional fugacity expansion with imaginary $\mu_B$ and $\mu_S$. The result for $P^{BS}_{21}$ is shown in figure~\ref{fig:PBS21}. This coefficient includes contributions from $N-\Lambda$ and $N-\Sigma$ scattering. The negative trend indecates the presence of an repulsive interaction that cannot be described with the addition of more resonances.

The ratios of various fluctuations can be used to express the cumulants of the Baryon number distribution. This offers an observable for comparisons with heavy ion collision measurements of the proton number distribution. At the current precision level this can only be a rough comparison. If the precision is increased in the future, other effects should be taken into account, like the continuum limit on the lattice side, or volume fluctuations and on equilibrium effects on the experimental side. 

\begin{figure}
 \centering
 \includegraphics[width=0.49\textwidth]{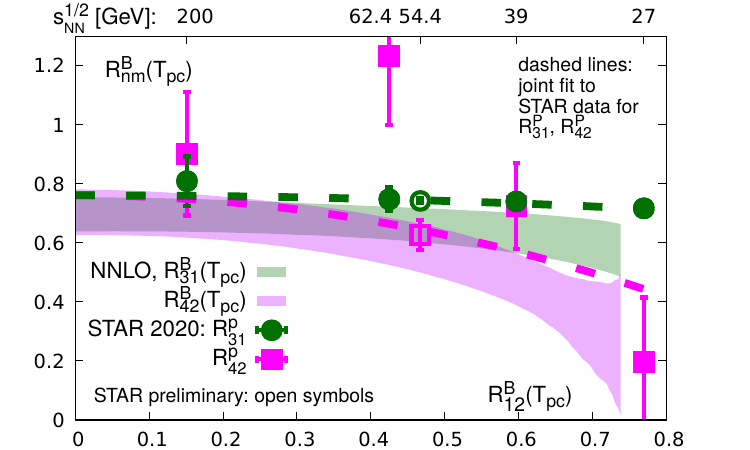}
 \hfill
 \includegraphics[width=0.49\textwidth]{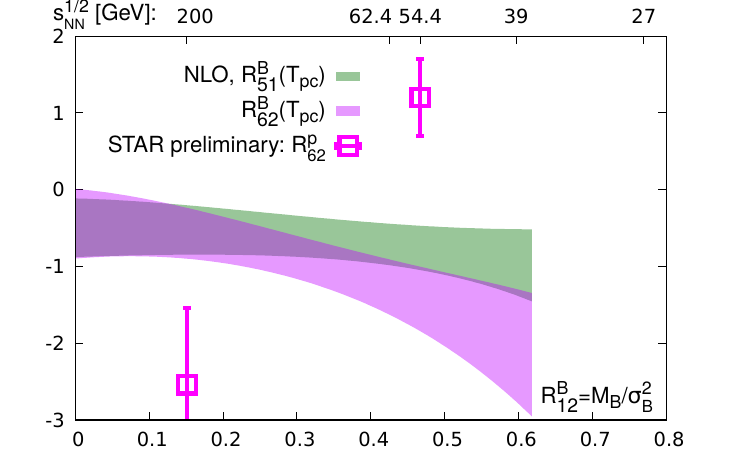}
 \caption{(Ref.~\cite{Bazavov:2020bjn}) Left: The ratios $R_{31}^B(T,\mu_B)= S_B \sigma_B^3/M_B$ and $R_{42}^B(T,\mu_B)= \kappa_B\sigma_B^2$ as a function of $R_{12}^B(T,\mu_B) = M_B/\sigma_B^2$ evaluated along the transition line in comparison to the data from the STAR collaboration (Ref.~\cite{Adam:2020unf,Nonaka:2020crv}).  Right: The ratios $R_{51}^B(T,\mu_B)$  and $R_{62}^B(T,\mu_B)$ as a function of
$R_{12}^B(T,\mu_B)$  in comparison to the data from the STAR collaboration (Ref.~\cite{Nonaka:2020crv}).}
 \label{fig:fluctuationsHotQCD}
\end{figure}

\begin{figure}
 \centering
 \includegraphics[width=0.6\textwidth]{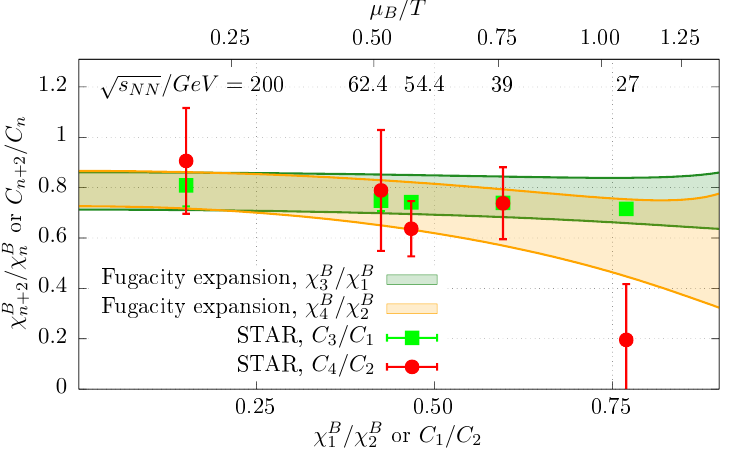}
 \caption{(Ref.~\cite{Bellwied:2021nrt}) The ratios $R_{31}^B(T,\mu_B)= S_B \sigma_B^3/M_B$ and $R_{42}^B(T,\mu_B)\equiv \kappa_B\sigma_B^2$ as a function of $R_{12}^B(T,\mu_B) = M_B/\sigma_B^2$ evaluated along the transition line in comparison to the data from the STAR collaboration (Ref.~\cite{Adam:2020unf}).}
 \label{fig:2DFluct}
\end{figure}

Figure~\ref{fig:fluctuationsHotQCD} from Ref.~\cite{Bazavov:2020bjn} shows the ratios
\begin{align}
 R_{31}^B(T,\mu_B)&= \frac{\chi^B_3}{\chi^B_1}= \frac{S_B \sigma_B^3}{M_B}\\
 R_{42}^B(T,\mu_B) &= \frac{\chi^B_4}{\chi^B_2}= \kappa_B\sigma_B^2\\
 R_{51}^B(T,\mu_B) &= \frac{\chi^B_5}{\chi^B_1}\\
 R_{62}^B(T,\mu_B) &= \frac{\chi^B_6}{\chi^B_2}
\end{align}
   as a function of $R_{12}^B(T,\mu_B) = \frac{M_B}{\sigma_B^2}$ evaluated along the transition line. Here $M_B$, $\sigma_B^2$, $S_B$ and $\kappa_B$ refer to the first four cumulants of the net baryon number distribution (mean, variance, skewness and kurtosis).  While the two lower order ratios are a  continuum estimate from $N_t=8$ and $N_t=12$ lattices, $R_{51}^B(T,\mu_B)$  and $R_{62}^B(T,\mu_B)$ are computed only on an $N_t=8$ lattice. A corresponding result from Ref.~\cite{Bellwied:2021nrt}, which has already been discussed above, is shown in figure~\ref{fig:2DFluct}.  

\begin{figure}
 \centering
  \includegraphics[width=0.33\textwidth]{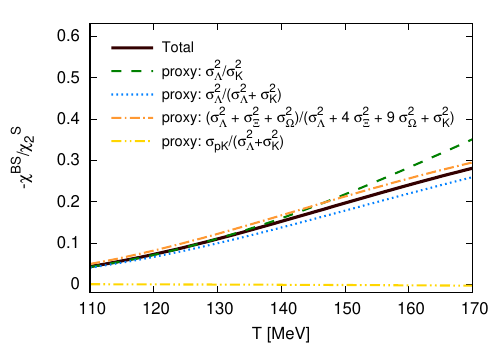}
 \hfill
  \includegraphics[width=0.31\textwidth]{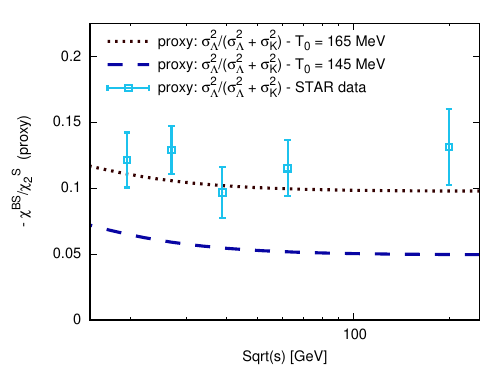}
\hfill
  \includegraphics[width=0.33\textwidth]{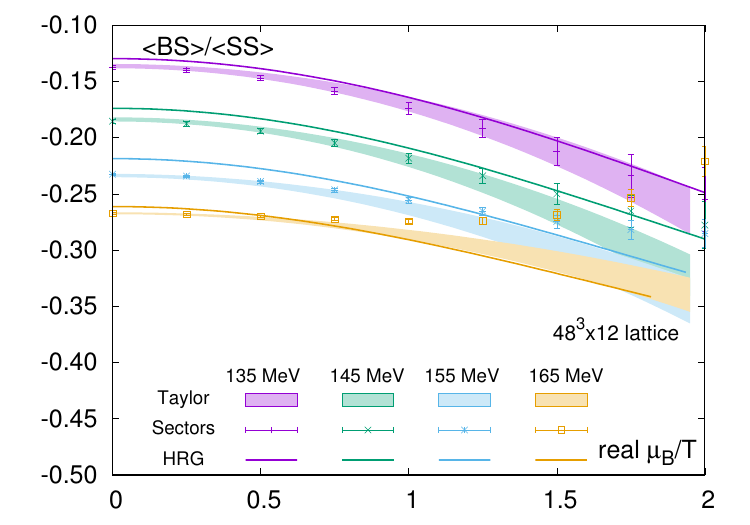}
  \caption{(Ref.~\cite{Bellwied:2019pxh}) Left: Calculations for different combinations of praticle number cumulants (proxis) that could be measurable in heavy ion collision expermients and the total of $-\frac{\chi^{BS}_{11}}{\chi^S_2}$ that is accessible in lattice QCD calculations  with the Hadron Resonance Gas (HRG) model. 
  Middle: Comparison of the proxy $\sigma^2_\Lambda/(\sigma^2_\Lambda+\sigma^2_K)$ 
  for two temperatures in the HRG model and experimental results from Ref.~\cite{STAR:2017tfy}
  Right: Comparison between the Taylor (see equation~(\ref{eqn:taylor})) and the sector (see equation~(\ref{eqn:fugacity})) method for $-\frac{\chi^{BS}_{11}}{\chi^S_2}$ on an $48^3\times12$ lattice, as well as results from the HRG model.}
  \label{fig:strangnessFluct}
\end{figure}
To include strange particles like $K$ or $\Lambda$ in the comparison with lattice calculations, suitable observables have to be constructed. Ref.~\cite{Bellwied:2019pxh} used the Hadron Resonance Gas (HRG) model to compare different proxies (see left of figure~\ref{fig:strangnessFluct}). The proxy $\sigma^2_\Lambda/(\sigma^2_\Lambda+\sigma^2_K)$ for the fluctuation ratio $-\frac{\chi^{BS}_{11}}{\chi^S_2}$ was further investigated both by comparing the HRG and the experimental (see middle of figure~\ref{fig:strangnessFluct}) results, as well as lattice calculations from both the Taylor and sector method (see right of figure~\ref{fig:strangnessFluct}).

\section{The equation of state}
\begin{figure}
 \centering
 \includegraphics[width=0.48\textwidth]{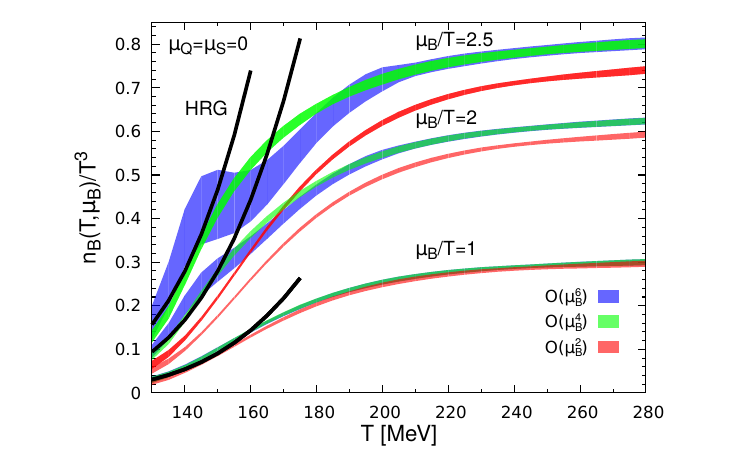}
 \hfill
 \includegraphics[width=0.48\textwidth]{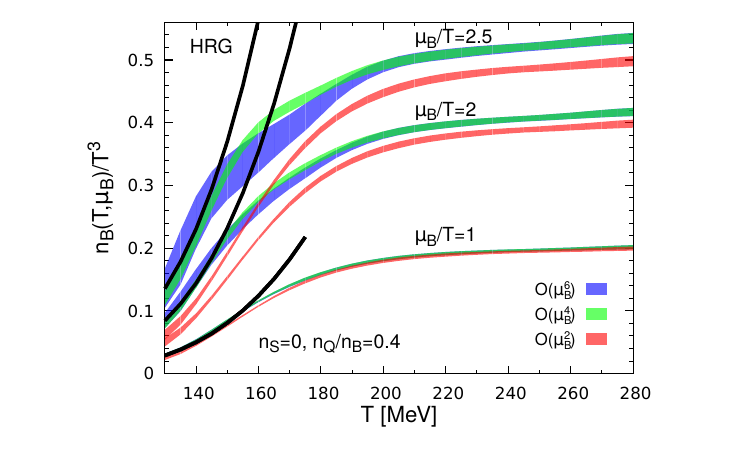}
 \caption{(Ref.~\cite{Bazavov:2017dus}) The extrapolation to finite baryon number $n_B$ done by the Taylor method (see equation~(\ref{eqn:taylor})): Left: $\mu_Q=\mu_S=0$, Right: $n_S=0$ and $n_Q/n_B=0.4$}
 \label{fig:EoSHotQCD}
\end{figure}

\begin{figure}
 \includegraphics[width=0.32\textwidth]{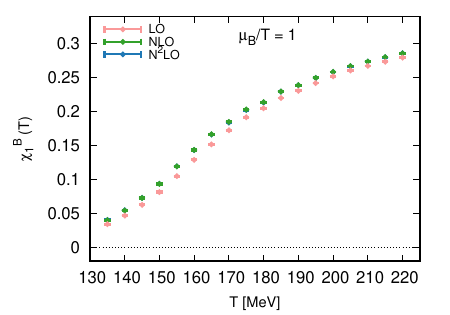}
 \hfill
 \includegraphics[width=0.32\textwidth]{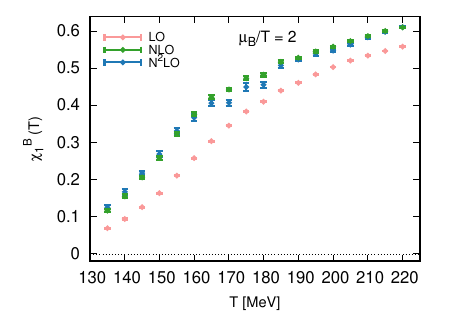}
 \hfill
 \includegraphics[width=0.32\textwidth]{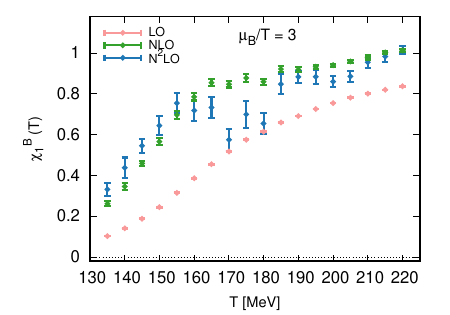}\\
 \includegraphics[width=0.32\textwidth]{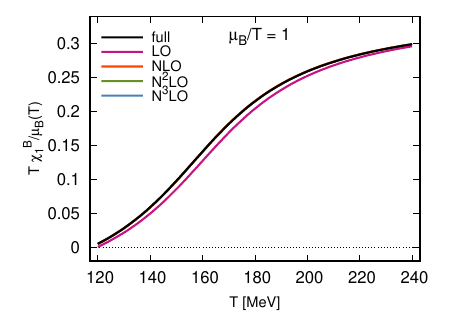}
 \hfill
 \includegraphics[width=0.32\textwidth]{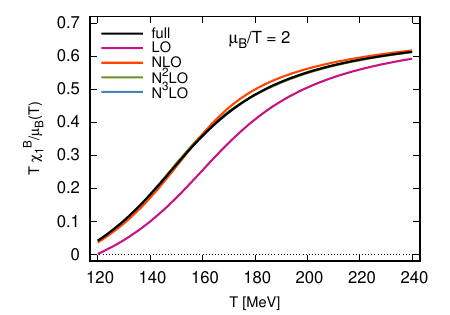}
 \hfill
 \includegraphics[width=0.32\textwidth]{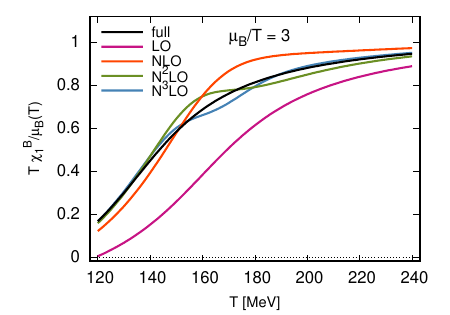}
 \caption{(Ref.~\cite{Borsanyi:2021sxv}) Top row: The extrapolation of $\chi^B_1$ with data from Ref.~\cite{Borsanyi:2018grb}) on $N_t=12$ lattices to different chemical potentials. Bottom row: The same extrapolation in a simple toymodel with different orders of the Taylor expansion.  }
 \label{fig:EoS}
\end{figure}

\begin{figure}
 \centering
  \includegraphics[width=0.49\textwidth]{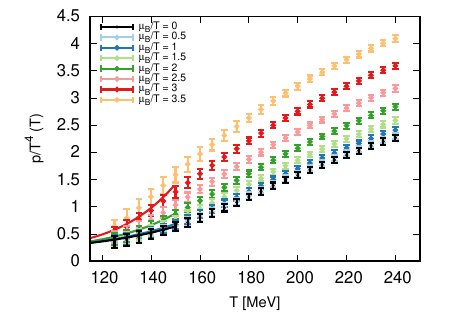}
  \hfill
  \includegraphics[width=0.49\textwidth]{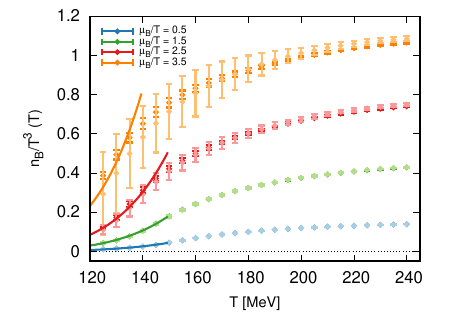}
  \caption{(Ref.~\cite{Borsanyi:2021sxv}) The pressure and the baryon number from the resummed extrapolation. The lighter bars show the increased error by the inclusion of a higher order term.}
  \label{fig:EoSNew}
\end{figure}

Another quantity that has been investigated for long time is the equation of state. In the following some progress on the baryon number $n_B$ will be discussed. When extrapolated to finite $\mu_B$ with a Taylor expansion up to $\mu_B^6$ it shows an increase in the error around the transition temperature, which leaves room for unexpected behavior.  (see figure~\ref{fig:EoSHotQCD} and top row of figure~\ref{fig:EoS}). This has been observed by different groups and on different data sets (Refs.~\cite{Bazavov:2017dus,Borsanyi:2021sxv}). In Ref.~\cite{Borsanyi:2021sxv} a simple toy model suggest that this behavior could linked to the cut-off in the Taylor series in this region, which is illustrated in the bottom row of figure~\ref{fig:EoS}. A possible explanation for the extra challenges around the transition temperature might be that the extrapolation has to cover both the hadronic and the quark gluon plasma phase. Therefore a new extrapolation scheme from imaginary $\mu_B$ was proposed that shows a smooth behaviour as can be seen in figure~\ref{fig:EoSNew}. Now the inclusion of an extra term does increase the error more broadly over all temperatures. This behavior can make the results more suitable, if the lattice results are taken as input in hydrodynamic models, where often only the central value is used.

\section{The critical endpoint}
The most sought after point in the QCD phase diagram is the critical endpoint, where the analytical transition between quark gluon plasma and hadrons becomes second order. However a comprehensive determination of the point is highly challenging. In this section I will briefly discus three different ansatzes that can lead to some inside despite all the challenges.
\subsection{Lee-Yang edge singularities}
\begin{figure}
 \centering
 \includegraphics[width=0.55\textwidth]{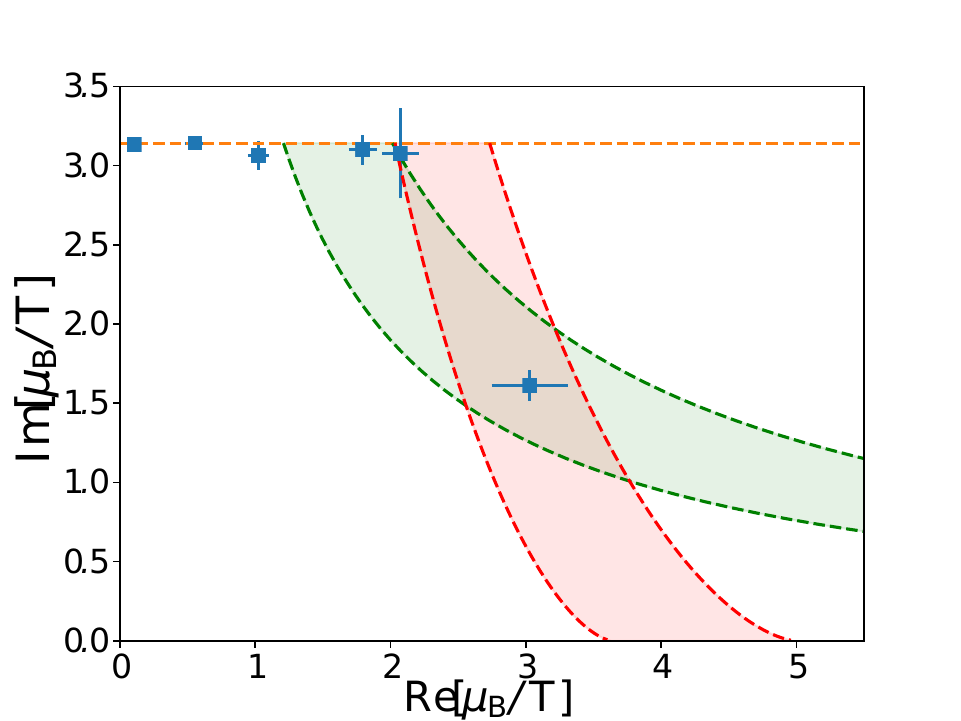}
 \hfill
 \includegraphics[width=0.35\textwidth]{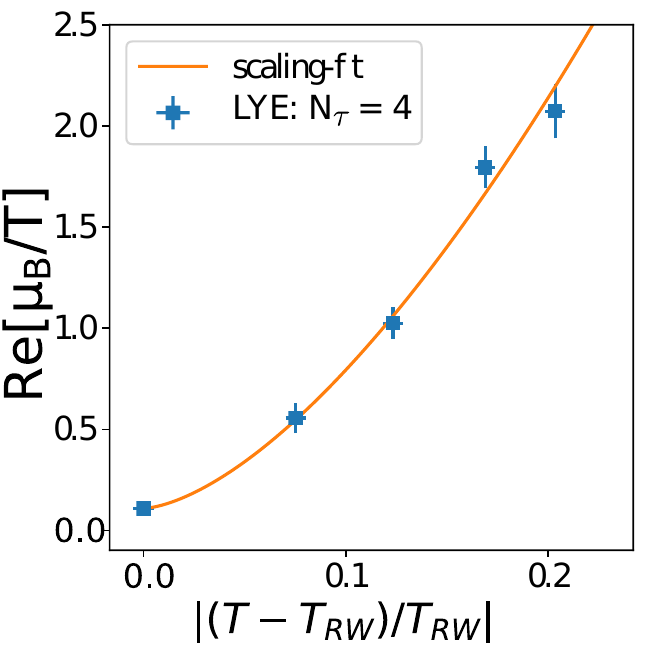}
 \caption{(Ref.~\cite{Nicotra:2021ijp,Dimopoulos:2021vrk}) Blue points: Results for Lee-Yang edge singularities from $N_t=4$ and 6 lattices. Left: Position of Lee-Yang edge singularities in the complex plain. Shaded areas show the expected ares for the Roberge-Weiss ($Z(2)$) in orange, the chiral ($O(4)/O(2)$) in green and the critical endpoint ($Z(2)$) in red. Right: The scaling fit of the Lee-Yang edge singularities associated with the Roberge-Weiss critical endpoint. The shaded areas are extracted from the expected scaling behavior that is stated for each critical endpoint.}
 \label{fig:LYE}
\end{figure}

 One way to search for a critical endpoint in the QCD phase diagram is to look for Lee-Yang edge singularities (Ref.~\cite{Lee:1952ig}). There have been recent efforts (Refs.~\cite{Giordano:2019gev,Giordano:2020roi,Dimopoulos:2021vrk}) to use reweighing to determine the leading singularities in the complex plain. New results presented at this years lattice conference use the fluctuations $\chi^B_1$, $\chi^B_2$ and $\chi^B_3$ to find a rational approximation for this quantities. This approximation can then be used to determine the Lee-Yang edge singularities by solving for the zeros of the polynomial in the denominator. The simulation were done with 2+1-flavors of HISQ quarks on lattices with $N_t=4$ and 6 and shown in figure~\ref{fig:LYE}. On the left side shaded areas are extracted from the expected scaling behavior of different critical endpoints: The first one is the Roberge-Weiss critical endpoint which is accessible from direct simulations at imaginary chemical potential. The second one is the chiral chritical endpoint which is discussed in section~\ref{sec:Columbia}. The third one is the QCD critical endpoint. Most singularities that could be determined seem to correspond to the Roberge-Weiss transition, for which the scaling behavior is shown on the right side of figure~\ref{fig:LYE}.

\subsection{Universality}
\begin{figure}
 \centering
 \includegraphics[width=0.4\textwidth]{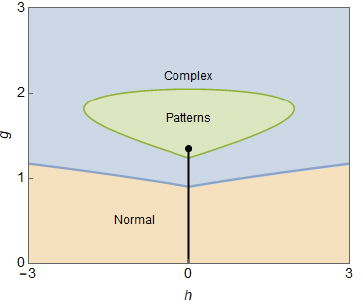}
 \hspace{1cm}
 \includegraphics[width=0.32\textwidth]{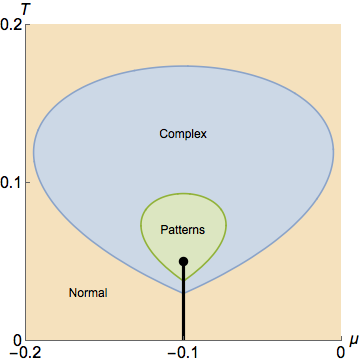}
 \caption{(Ref.~\cite{Schindler:2021otf,Schindler:2021cke}) Patterns around a critical endpoint in a QCD-inspired heavy quark model.}
 \label{fig:universality}
\end{figure}

An important feature of a second order phase transition is the universality in its vicinity. This would allow to extract information about the QCD phase diagram by studying different theories or models, which can be more accessible compared to full QCD at finite density. To make this possible the universality class and a mapping between QCD and the theory under investigation has to be established. Results of one possible model were presented at this years lattice conference (Refs.~\cite{Schindler:2021psi,Schindler:2021cke}). In Ref.~\cite{Schindler:2021otf} a $\mathcal{PT}$ symmetric quark model with $Z(2)$ symmetry was studied. Its action is given as
\begin{equation}
  S(\phi,\chi) = \sum_x \frac{1}{2}(\nabla_\mu \phi)^2 + \frac{1}{2}(\nabla_\mu \chi)^2 + V(\phi,\chi)
\end{equation}
with
\begin{equation}
 V(\phi,\chi) =  \frac{1}{2}m_\chi^2 \chi^2   -ig\phi\chi+ U(\phi)+h\phi.
\end{equation}
Around the chritical endpoint patterend regions are found, which the authors compare to the patterns of nuclear pasta (see e.g. Ref.~\cite{Caplan:2016uvu}). In figure~\ref{fig:universality} the phase diagram for this model is shown.

\subsection{Functional methods}
\begin{figure}
 \centering
 \includegraphics[width=0.5\textwidth]{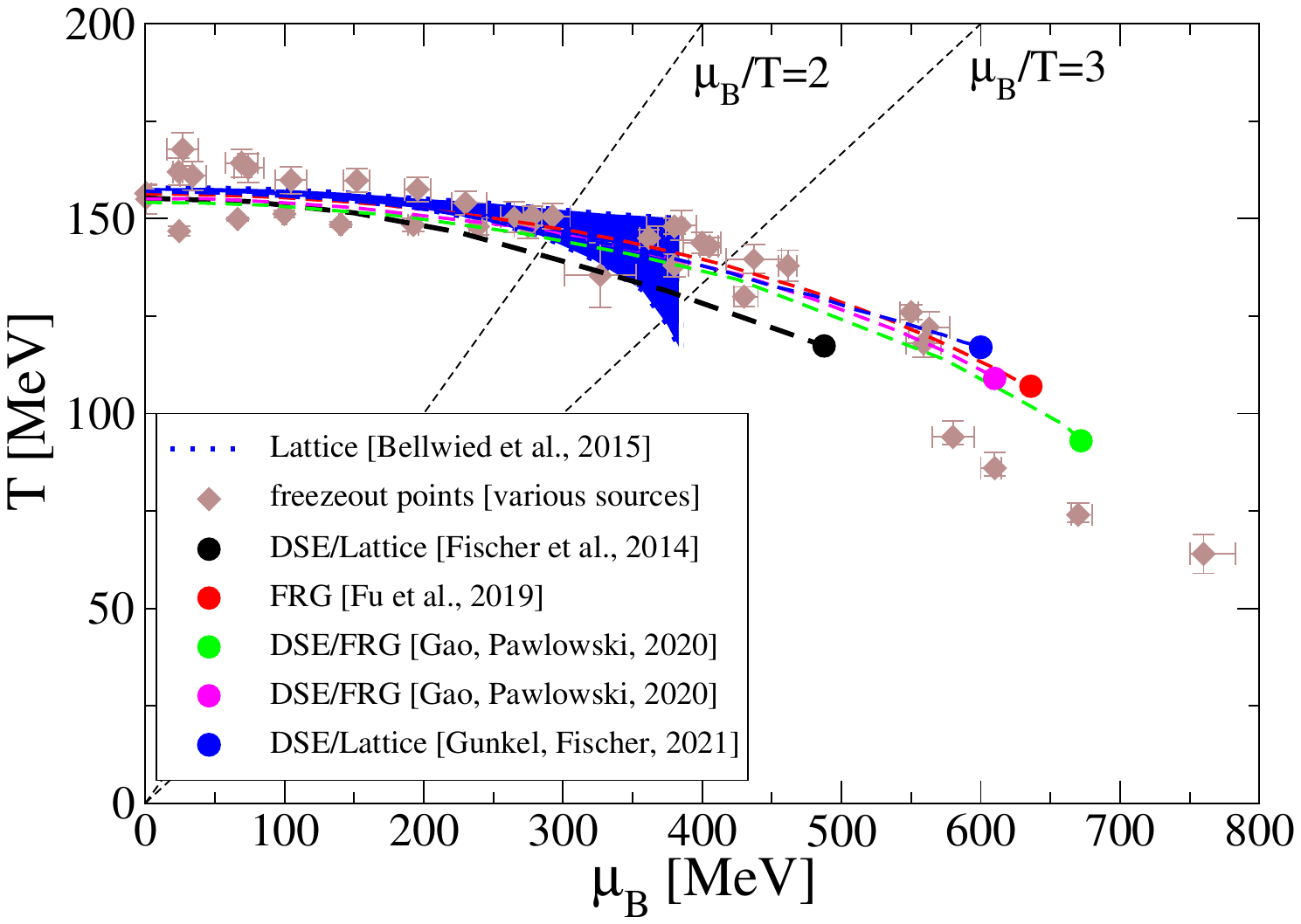}
 
 \caption{(Ref.~\cite{TalkDSE}) Overview over results on the QCD transition line from Lattice QCD, FRG and DSE (Refs.~\cite{Bellwied:2015rza,Fu:2019hdw,Gunkel:2020wcl,Fischer:2014mda,Gao:2020fbl,Gao:2020qsj}). as well as results on the freeze-out line from heavy ion collision experiment.  }
 \label{fig:DSE}
\end{figure}

Another approach, beside lattice QCD, to study QCD from its fundamental degrees of freedom are the functional methods. Under this name, both Functional Renormalization group (FRG) (see e.g. Refs.~\cite{Fu:2018qsk, Fu:2019hdw, Leonhardt:2019fua, Braun:2019aow, Braun:2020ada, Dupuis:2020fhh}) and Dyson-Schwinger equations (DSE) (see e.g. Refs.~\cite{Roberts:2000aa, Fischer:2014ata, Gao:2016qkh, Gao:2020qsj, Fischer:2018sdj, Isserstedt:2019pgx, Gao:2020qsj}) are combined. DSE rely on an infinite tower of equations that give an exact representation of QCD. However to solve those equations this tower has to be truncated, which introduces an error that cannot be estimated at the moment. The FRG method relies on a functional integro-differential equation to describe the flow of the action (Ref.~\cite{Morris:1993qb}). Again a truncation introduces an unknown error.
An idea for an error estimation can be gained from the comparison between different truncations, methods or with lattice and experimental results. An overview over some recent determinations of the crossover line in the QCD phase diagram and the critical endpoint in comparison with lattice and freeze-out results can be seen in figure~\ref{fig:DSE}. The newer computations agree well with the lattice results and the critical endpoint estimations seem to be concentrated in an relative small part of the phase diagram. However further investigation and comparison of other observables is necessary before any conclusions can be drawn.

\clearpage
\section{Conclusion}
As was evident as this conference, the QCD phase diagram is a very active topic of research with lots of exciting challenges for lattice QCD. The understanding of the phase diagram is a fascinating subject for theory and experiments. In addition, with new experimental results expected in the next years, further progress in theory is needed to be able to give input and enable comparisons as close to the experimental conditions as possible. This proceedings tried to update and add on last year's review (Ref.~\cite{Guenther:2020jwe}) with a focus on activities at the conference. Still, not all aspects could be discussed. For example, another axis of the phase diagram, for which new results are available, is the dependence on the isospin chemical potential (Refs.~\cite{Chabane:2021pfk,Cuteri:2021nnf,Brandt:2021yhc}).
For many aspects the next years promise to be interesting, with the hope of new insights and understanding in various areas.

\acknowledgments
The project leading to this publication has received funding from Excellence Initiative of Aix-Marseille University - A*MIDEX, a French ``Investissements d'Avenir'' programme, AMX-18-ACE-005.
Several of the projects discussed here also received support from the BMBF Grant No. 05P18PXFCA.
This proceeding is part of a project that has received funding from the European Union's Horizon 2020 research and innovation program under grant agreement STRONG -- 2020 - No 824093.
The author gratefully acknowledge the Gauss Centre for Supercomputing e.V.  (www.gauss-centre.eu) for
funding several projects, which are part of this review, by providing computing time on the GCS Supercomputer
HAWK at HLRS, Stuttgart.
The author thanks Lukas Varnhorst for proofreading and discussion.

\bibliographystyle{JHEP}
\bibliography{finiteT}

\end{document}